\documentclass[12pt]{scrartcl}
\usepackage[affil-it,auth-lg]{authblk}

\usepackage[latin1]{inputenc}
\usepackage{amsmath}
\usepackage{amsfonts}
\usepackage{amssymb}
\usepackage{bm,amsmath,amssymb,amsfonts,latexsym,psfrag}
\usepackage{bm}
\usepackage{amssymb}
 \usepackage{array}
\usepackage{mathtools}
\usepackage{amsthm}
\usepackage{newlfont}
\usepackage{esdiff}
\usepackage{multirow}

\usepackage{graphicx}
\usepackage{parskip}
\usepackage{graphicx}
\usepackage{url}                
\usepackage{bm,amsmath,amssymb,amsfonts,latexsym,psfrag}

\renewcommand{\d}{\mathrm{d}}

\newcommand{\U}{ {\cal U}}

\usepackage[small,it]{caption}
 \usepackage{amssymb}
 \usepackage{enumerate}
 \usepackage{color}
 \let\origthanks\thanks\renewcommand\thanks[1]{\begingroup\let\rlap\relax\origthanks{#1}\endgroup}

\titlehead{
}
\title{Unparticle Casimir Effect}

\author[a,b]{Antonia~M.~Frassino\footnote{\textit{E-mail:} \texttt{frassino@fias.uni-frankfurt.de}}}
\author[a,b]{Piero Nicolini\footnote{\textit{E-mail:} \texttt{nicolini@fias.uni-frankfurt.de}}}
\author[c]{Orlando Panella\footnote{\textit{E-mail:} \texttt{orlando.panella@pg.infn.it}}}

\affil[a]{Frankfurt Institute for Advanced Studies, Ruth-Moufang-Strasse 1, D-60438 Frankfurt am Main, Germany}
\affil[b]{Institut f\"{u}r Theoretische Physik, J. W. Goethe-Universit\"{a}t,  Max-von-Laue-Strasse 1, D-60438 Frankfurt am Main, Germany}
\affil[c]{Istituto Nazionale di Fisica Nucleare, Sezione di Perugia, Via A.~Pascoli, I-06123 Perugia, Italy}

\date{\large\today}

\begin{document}
\maketitle
\begin{abstract}
In this paper we present the un-Casimir effect, namely the study of the Casimir energy in the presence of an unparticle component in addition to the electromagnetic field contribution.  
The distinctive feature of the un-Casimir effect is a fractalization of metallic plates. This result emerges through a new dependence of the Casimir energy on the plate separation that scales with a continuous power controlled by the unparticle dimension.  
As long as the perfect conductor approximation is valid, we find  bounds on the unparticle scale that are independent of the effective coupling constant between the scale invariant sector and ordinary matter. 
We find regions of the parameter space such that for plate distances around  $5\mu$m and larger the un-Casimir bound wins over the other bounds.
\end{abstract} 


 
\section{Introduction}
Recently, a massive extension to the Standard Model was proposed in which scale invariance is preserved, provided these new particles are weakly interacting and appear in non-integer numbers~\cite{PhysRevLett.98.221601}.  The topic has intersected a huge variety of fields, spanning astrophysics  neutrino physics, AdS/CFT duality and quantum gravity.

Scale invariance for massive fields  can be described through Banks-Zacks ($\cal{BZ}$) fields~\cite{Banks:1981nn}. At some very high energy scale $M_{\mathcal{U}}$, the Standard Model 
fields interact with a sector exhibiting a non-trivial infrared $\cal{BZ}$ fixed point, ${\cal L}_{\mathrm{int}}={(M_{\mathcal{U}})^{-k}}{\cal O}_{\mathrm{SM}}{\cal O}_{\cal{BZ}}$,
where the field operators {must} have dimensions $d_{\mathrm{SM}}$, $d_{\cal{BZ}}$ and
$k=d_{\mathrm{SM}}+d_{\cal{BZ}}-4$. 
{Since} the Banks-Zaks fields {are not observed in nature}, their suppression requires that the scale $M_{\mathcal{U}}$ is somewhere between current experimentally-accessible scales and the Planck scale. 

At a {second energy} scale $\Lambda_{\mathcal{U}} <  M_{\mathcal{U}}$, the $\cal{BZ}$ sector {develops} scale-invariant properties and the particle number is controlled by a continuous parameter $d_{\mathcal{U}}\neq d_{\cal{BZ}}$. This is equivalent to saying that $\cal{BZ}$ fields undergo a dimensional transmutation to become unparticles via
\begin{equation}
\label{interaction1}
\frac{1}{(M_{\mathcal{U}})^k}{\cal O}_{\mathrm{SM}}{\cal O}_{\cal{BZ}}\xrightarrow[\sim  \Lambda_{\mathcal{U}}]{}\ \lambda \ \frac{1}{(\Lambda_{\mathcal{U}})^{k_\mathcal{U}}}
{\cal O}_{\mathrm{SM}}{\cal O}_{\mathcal{U}}
\end{equation}
where the unparticle operator ${\cal O}_{\mathcal{U}}$ has dimension $d_{\mathcal{U}}$ and $k_\mathcal{U}=d_{\mathcal{U}}+d_{\mathrm{SM}}-4$. 
We note that the resulting interaction term depends on a dimensionless coupling constant 
$
\lambda=\left(\Lambda_{\mathcal{U}}/M_{\mathcal{U}}\right)^k < 1.
$
{Unitarity constraints from} conformal field theory (CFT) {necessitate a} lower bound { on the unparticle dimension} $d_{\mathcal{U}} \ge 1$~\cite{Grinstein2008367}. Normally only operators with $d_{\mathcal{U}} \le 2$ are considered because for $d_{\mathcal{U}} \ge 2$ the calculations become sensitive to the ultraviolet sector  and therefore less predictive. For a discussion about a more general form of these operators see for instance the HEIDI models \cite{Bij1,Bij2,Bij3,Bij4}.

The phenomenology of unparticles and the associated signals at high energy colliders (LEP, LHC) have received  attention in recent literature~\cite{Cheung:2009zz,Cheung:2008xu,PhysRevD.84.015010,CMS-PAS-EXO-11-043,CMS-PAS-EXO-09-011,Aliev:2017bme}.
 
On the theoretical side, bounds on the parameter space have been derived by computing the unparticle contribution to the muon anomaly \cite{PhysRevLett.99.051803}.  
Unparticles also { provide} a relevant short scale modification to {gravitational interactions}.   Black hole solutions have been derived for the case of scalar \cite{Mureika:2007nc,Mureika:2008dx,Gaete:2010sp} and vector \cite{Mureika:2010je}  unparticle exchange.  A significant cha\-ra\-cteristic of these solutions is the fractalization of the event horizon, whose dimension is {a function of} the unparticle parameter $d_{\cal U}$. This feature has been confirmed by subsequent studies of the spectral dimension, as an indicator of the short-scale spacetime dimension perceived by an unparticle probe \cite{Nicolini2011290}.  In addition, such unparticle modifications to gravity offer compelling effects that can be {observed through short-scale  deviations} to Newton's law \cite{Goldberg:2007tt} and on the energy levels of the hydrogen atom \cite{Wondrak:2016itb}.

In this paper, we will analyze the Casimir effect in the presence of a weakly-coupled unparticle sector, which we will refer to as the \emph{un-Casimir} effect. The Casimir effect has been discussed within various scenarios beyond the Standard Model including compactified extra dimensions~\cite{Poppenhaeger:2003es} and minimal length theories~\cite{Panella:2007kd,Frassino:2011aa}.   Using the un-Casimir effect, we present a privileged testbed for setting bounds on relevant regions of the parameter space $[\Lambda_{\cal{U}},d_{\cal{U}}]$ governing  unparticle physics. The un-Casimir effect also offers an intriguing phenomenon of plate fractalization in agreement with the above discussion. 

\section{Unparticle contribution to the Casimir effect}
\label{uncasimir}

In line with \eqref{interaction1} we assume the existence of an unparticle vector field ${\cal A}^{\cal U}_\mu$ of scaling dimension $d_U$ which couples with the standard model electron Dirac current $J^\mu=\bar{\psi}\gamma^\mu \psi$ via the interaction:
\begin{equation}
\label{basic-interaction}
{\cal L_{\mathrm{int}}} = \frac{\lambda}{\Lambda_{\cal U}^{d_{\cal U}-1}} {\cal A}^{\cal U}_\mu \, \bar{\psi}\gamma^\mu \psi.
\end{equation} 
Interactions  of the form given  in \eqref{basic-interaction} have been extensively used in the literature in order to study the phe\-no\-me\-no\-logy of unparticles~\cite{Georgi2007275,Liao:2007bx,Chen:2007aa}.
In the presence of  perfectly conducting parallel plates at a distance $a$, the interaction in \eqref{basic-interaction} will be responsible for  a Casimir effect for the field ${\cal A}_{\cal U}^\mu$ in much the same way the interaction ${\cal L_{\mathrm{int}}} = {e}\,{\cal A}_\mu \,\bar{\psi}\gamma^\mu \psi$ implies the Casimir effect in QED.
For ease of presentation, in the following section 
we discuss the Casimir effect mediated by a \emph{scalar} unparticle field using the 
scalar field analogy \footnote{Alternatively one can consider the Lagrangian $$
{\cal L_{\mathrm{int}}} = \frac{\lambda}{\Lambda_{\cal U}^{d_{\cal U}-1}} J_{\cal U}\phi_{\cal U},$$
that describes the coupling between electrons and scalar un-particles. Such interaction has to vanish in the limit $d_{\cal U}\to 1$ but it is legitimate for $d_{\cal U}\neq 1$.}.  {This is routinely} done in QED, where the actual result is just twice that of a scalar field due to the two physical photon polarisations. For the case of un-particles such a working hypothesis remains valid, since the energy of longitudinal modes related to superposition of massive fields (see \eqref{unprop} below) do not exceed $1 \%$ of that of the transverse polarizations.   A full treatment of the Casimir effect mediated by a vector unparticle field will be presented in \cite{FNP16}. 
  
The Casimir energy~\cite{casimirHB} is often described by the shift in the sum of the zero point energies  of the normal modes of the electromagnetic field induced by  geometrical boundary conditions. 
Such a Casimir energy  can be written by means of the density of states $dN/d\omega$  which in quantum field theory (QFT)  is related to the imaginary part of the trace (over both space and spinor degrees of freedom of the field under consideration) of the Feynman propagator~\cite{A.-A.-Abrikosov:1965qy,PhysRevD.72.021301}.  It has been shown in \cite[p. 47]{bordag2014advances} that for a generic massive, scalar field the vacuum energy can be written as
\begin{equation} 
{\cal E}_{{\mathrm{vac}}}(m)=i\int \d^3x\ \partial_0^2\, \left. \text{\sf{\bf{D}}}(x,x';m^2)\right|_{x=x'}.
\label{eq:casen}
\end{equation}
Here $\text{\sf{\bf{D}}}(x,x';m^2)$ denotes  the Green's function of the massive scalar field.
Eq. \eqref{eq:casen} is obtained from the spatial integral of the vacuum expectation value of the
energy-momentum tensor
\begin{equation}
\langle 0|T_{00}|0\rangle=-\frac{i}{2}\left(\sum_{\mu=0}^3\partial_\mu\partial_\mu^\prime+m^2\right)\left. \,\text{\sf{\bf{D}}}(x,x';m^2)\right|_{x=x'}.
\end{equation}

In case of an unparticle field, one has \cite{PhysRevLett.98.221601} a mo\-di\-fied Feynman propagator~\cite{Nicolini2011290,Georgi2007275,Gaete2008319} given by the following representation:
\begin{eqnarray}
\text{\sf{\bf{D}}}_{\cal U}(x,x')&=&\frac{A_{d_{\cal U}}}{2\pi (\Lambda_{\cal U}^2)^{d_{\cal U}-1}}\, \int_0^\infty dm^2 (m^2)^{d_{\cal U}-2} \, \text{\sf{\bf{D}}}(x,x';m^2)\nonumber \\
A_{d_{\cal U}}& =& \frac{16\, \pi^{5/2}}{(2\pi)^{2d_{\cal U}} } \, \frac{\Gamma(d_{\cal U}+1/2)}{\Gamma(d_{\cal U}-1)\, \Gamma(2d_{\cal U})}
\label{unprop}
\end{eqnarray} 
\textit{i.e.} it is a linear continuous superposition of Feynman propagators of fixed mass $m$.  
When the conformal dimension tends to unity  ($d_{\cal U} \to 1 $) the unparticle propagator reduces to that of an ordinary massless  field $D_{\cal U}(p^2) \to1/p^2$~\cite{Georgi2007275}.
The above propagator can be expressed in terms of the unparticle generating functional $Z_{\cal U}[J]$ \cite{Gaete2008319,Gaete:2014rwa}. The net result is 
\begin{equation}
\text{\sf{\bf{D}}}_\U\left(\, x, x' \,\right)=\hat{F}_\U^{-1}(\Box)\  \text{\sf{\bf{D}}}\left(\, x, x'\,\right)
\label{eq:greenrelation}
\end{equation}
where $\text{\sf{\bf{D}}}\left(\, x, x'\,\right)$ is the Green's function for massless scalars and $\hat{F}_\U^{-1}(\Box)$ is a non-local operator defined as \cite{Gaete2008319,Gaete:2014rwa}
\begin{equation}
\hat{F}_\U (\Box)\equiv \frac{2 \, \sin\left(\, \pi\, d_U\,\right)\, 
}{A_{d_U}} \,
\left(\, \frac{-\Box}{\Lambda^2_\U}\,\right)^{1-d_\U}.
\end{equation}
Eq. \eqref{eq:greenrelation} shows that the unparticle propagator actually solves the massless Green function equation, since $[\, \Box , \, \hat{F}_\U (\Box) \,]=0$. This fact allows us to assume 
\begin{equation}
\langle0|T_{00}^\U|0\rangle\equiv \left[\left.-\frac{i}{2}\left(\sum_{\mu=0}^3\partial_\mu\partial_\mu^\prime\right)\, \text{\sf{\bf{D}}}_\U(x,x^\prime)\right|_{x=x'}\right].
\label{eq:utimunu3}
\end{equation}
The above assumption turns out to be in complete agreement with previous results
for the non-local quantum stress tensor in a variety of contexts (e.g. for a generic
operator $\hat{F}_\U (\Box)$). See for instance \cite{Gaete:2010sp} for the scalar un-particle mediated gravity  and \cite{Barvinsky:2003kg,Modesto:2010uh,Modesto:2011kw} for other non-local deformations where it is found:
%
\begin{equation}
\langle0|T_{00}^\U|0\rangle =\hat{F}^{-1}_\U(\Box)\,  \langle0|T_{00}|0\rangle,
\label{eq:utimunu2}
\end{equation}
with $\langle0|T_{00}|0\rangle$ referring to a massless scalar field. Indeed by using \eqref{eq:greenrelation} in \eqref{eq:utimunu2} one obtains \eqref{eq:utimunu3}.

By  inserting  \eqref{unprop} in \eqref{eq:utimunu3}  one  has that the unparticle Casimir energy $\cal{E}_{\cal U}^{\mathrm{C}}$ reads:
\begin{equation}
\label{central}
\cal{E}_{\cal U}^{\mathrm{C}} =\frac{A_{d_{\cal U}}}{\pi (\Lambda_{\cal U}^2)^{d_{\cal U}-1}}\,\int_0^\infty dm \, m^{2d_{\cal U}-3}\, \, \cal{E}^{\mathrm{C}}(m)\,  .
\end{equation}
Here we used the fact the derivatives $\partial_\mu\partial_\mu^\prime$ and the integration on $m$ commute. To derive the above result for the Casimir energy we apply in \eqref{eq:casen} geometric boundary conditions at the plates and subtract the free vacuum (no boundary) contribution.

The above formula shows that $\cal{E}_{\cal U}^{\mathrm{C}}$ can be  related  to  the standard Casimir energy of a scalar field of fixed mass $m$, \textit{i.e.},  $\cal{E}^{\mathrm{C}}(m)$.
We also note that \eqref{central} is formally equivalent to the Casimir effect in Randall-Sundrum type II models~\cite{PhysRevLett.83.4690,Randall199979}, where the hidden 3-brane taken to infinity generates a continuous spectrum of Kaluza-Klein excitations \cite{PhysRevD.76.015008}.

\section{Un-Casimir effect}
\label{scalarUNC}

To calculate ${\cal E}_{\cal U}^{\mathrm{C}}$ we consider the Casimir energy for a massive scalar field~\cite{Barton:1984kx,Barton1985231,Farina1997}:
\begin{equation}
\cal{E}^{\mathrm{C}}(m)= -\frac{1}{8\pi^2} \, \frac{m^2}{a}
 \sum_{n=1}^{\infty} \frac{1}{n^2} \, K_2(2 a m n),
 \label{ECmassm}
 \end{equation}
where $K_2(z)$ is a modified Bessel function of the second type.  
By inserting \eqref{ECmassm}  in \eqref{central}
we analytically find the unparticle Casimir energy that reads:
\begin{equation}
\label{finalscalar}
\cal{E}_{\cal U}^\mathrm{C} (a) = - \frac{1}{a^3} \, \frac{d_{\cal U} \, \zeta(2+2d_{\cal U})}{ (4\pi)^{2d_{\cal U}}}\, \frac{1}{(a\Lambda_{\cal U})^{2d_{\cal U}-2}}\, ,
 \end{equation}
 where  $\zeta(s) =\sum_{n=1}^{\infty} n^{-s}$  is the Riemann Zeta function.
In the limit $d_{\cal U} \to 1$,
\eqref{finalscalar}  reproduces the ordinary result for the Casimir effect  of the scalar massless field.
Despite the similarities with the Casimir effect of RSII type  \cite{PhysRevD.76.015008},  the final explicit result \eqref{finalscalar} carries important differences. The unparticle contribution $\cal{E}_{\cal U}^{\mathrm{C}}$  depends not only on the new energy scale but, more importantly, also on the conformal dimension $d_{\cal U}$. This makes the unparticle contribution sizable for  $d_{\cal U}$ approaching $1$. Comparatively, the RSII Casimir energy is always suppressed since it scales  as $(\kappa a)^{-1}\approx 10^{-28}$, where $\kappa\sim 10^{19}$ GeV is the curvature parameter of the warped dimension and the separation length is typically $a\sim 1$ $\mu$m.

As a related remark, we note that unparticles introduce a new and distinctive effect, \textit{i.e.}, a fractalization of metallic plates. This is evident by writing (\ref{finalscalar}) as
 \begin{equation}
 \cal{E}_{\cal U}^{\mathrm{C}} (a) = - \frac{1}{a^{\mathbb{D}+1}} \, \frac{d_{\cal U} \, \zeta(2+2d_{\cal U})}{(4\pi)^{2d_{\cal U}}}\, \frac{1}{(\Lambda_{\cal U})^{2d_{\cal U}-2}}\,.
 \end{equation}
 In the conventional case $d_{\cal U}=1$, one finds $\mathbb{D}=D$, corresponding to the topological dimension $D=2$ of the boundary (we recall that, on dimensional grounds, $\cal{E}^C (a)$  is an energy per unit {area}).  On the other hand, when $d_{\cal U}\neq 1$, the dimensional parameter $\mathbb{D}$ departs from integer values, a typical feature of fractal surfaces. Specifically one finds that $\mathbb{D}$ is completely {determined} by the unparticle dimension as $\mathbb{D}=2d_{\cal U}$. This result is in agreement with an equivalent fractalization of a black hole horizon obtained from scalar~\cite{Goldberg:2007tt,Mureika:2007nc,Gaete:2010sp} and vector~\cite{Mureika:2010je} unparticle exchange.  The fractality encoded in unparticles has also been studied from a more general viewpoint. Fractals require the introduction of dimensional probes like the spectral dimension, \textit{i.e.}, the dimension perceived by a diffusive process or random walker. As such, it has been shown that the complete fractalization of plates, \textit{i.e.}, $\mathbb{D}=2d_{\cal U}$ is a general result deriving from the spectral dimension for an unparticle field propagating on a manifold with topological dimension $D=2$~\cite{Nicolini2011290}.

Finally we find that, for two parallel metallic plates separated by a distance $a$, the total attractive energy reads
 \begin{equation}
 \label{uncasimirtotal}
 \cal{E}^{\mathrm{C}} (a)= - \frac{\pi^2}{720 a^3} \left[1+ \frac{720\, d_{\cal U} \, \zeta(2+2d_{\cal U})}{\pi^2 (4\pi)^{2d_{\cal U}}}\, \frac{1}{(a\Lambda_{\cal U})^{2d_{\cal U}-2}}\right].
 \end{equation}
The above result exhibits an additional contribution to the standard electromagnetic Casimir effect ($-\frac{\pi^2}{720a^3}$ or  twice that of the scalar case). 
Eq. (\ref{uncasimirtotal}) gives the definition of the spectral dimension of plates in terms of the Casimir energy as
$\mathbb{D}=-\frac{\partial\log\cal{E}^{\mathrm{C}} (a)}{\partial \log a}-1$,
where $\cal{E}^{\mathrm{C}} (a)$ and $a$ play the role of the return probability and the diffusion time respectively.
By  using (\ref{uncasimirtotal}) one finds
\begin{equation}
\mathbb{D}=\frac{2+(2d_{\cal U})\beta}{1+\beta},
\label{specplate}
\end{equation}
where  $\beta=\frac{720\, d_{\cal U} \, \zeta(2+2d_{\cal U})}{\pi^2 (4\pi)^{2d_{\cal U}}}\, \frac{1}{(a\Lambda_{\cal U})^{2d_{\cal U}-2}}$.
This formula shows, for $d_{\cal U}>1$, a dimensional flow interpolating the following two regimes. For large plate separation $a\gg 1/\Lambda_{\cal U}$ we recover the usual topological result, \textit{i.e.}, $\mathbb{D}\to 2$.  On the other hand in the unparticle dominated case $a\ll 1/\Lambda_{\cal U}$, plate fractalization takes place, \textit{i.e.}, $\mathbb{D}\to 2d_{\cal U}$. The conventional Casimir result, $\mathbb{D}=2$,  is recovered by taking the limit of (\ref{specplate}) for $d_{\cal U}\to 1$. 
As expected  $\Lambda_{\cal U}$ is the critical scale at which the transition between the two phases (ordinary matter and unparticles) occurs.

\section{Discussion}
The un-Casimir effect offers important phenomenological predictions. We can get an estimate of the unparticle scale $\Lambda_{\cal U}$ as follows. 
 If $\Delta_{\text{Cas}}$ is the relative error of the experimental measurement, by imposing that $\left|\left[\cal{E}^{\mathrm{C}} (a) -\cal{E}^{\mathrm{C}}_{\text{QED}} (a)\right]/\cal{E}^{\mathrm{C}}_{\text{QED}} (a)\right| \le \Delta_{\text{Cas}}$ 
 we obtain (for $d_{\cal U}\ne 1$) the bound on $\Lambda_{\cal U}$:
 \begin{equation}
 \Lambda_{\cal U} \geq \Lambda_a \equiv\frac{1}{a}\, \left[ \frac{720\, d_{\cal U} \, \zeta(2+2d_{\cal U})}{\pi^2 (4\pi)^{2d_{\cal U}}}\, \frac{1}{\Delta_{\text{Cas}}}\right]^{\frac{1}{2d_{\cal U} -2}}\, .
 \label{bound}
 \end{equation}
We notice that there is a strong dependence on the parameter $d_{\cal U}$. In particular for values of $d_{\cal U} $ slightly above 1 the bound on $\Lambda_{\cal U}$ is very strong while as soon as $d_{\cal U} $ increases the bound exponentially decreases.
\begin{figure}[t!]
\begin{center}

\includegraphics[width=0.46\textwidth]{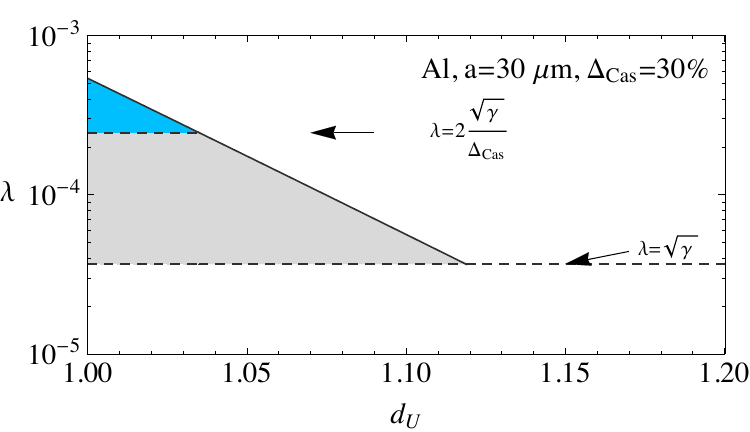} 
\caption{Contour plot (strong coupling regime) of  the ratio $\Lambda_{a}/\Lambda_{{\mu}}=1$. The region below the solid line corresponds to  $\Lambda_{a}/\Lambda_{{\mu}}>1$ (unCasimir provides the strongest bound).
\label{figex2} 
}
\end{center}

\end{figure}

\begin{figure}[t!]
\begin{center}

\includegraphics[width=0.46\textwidth]{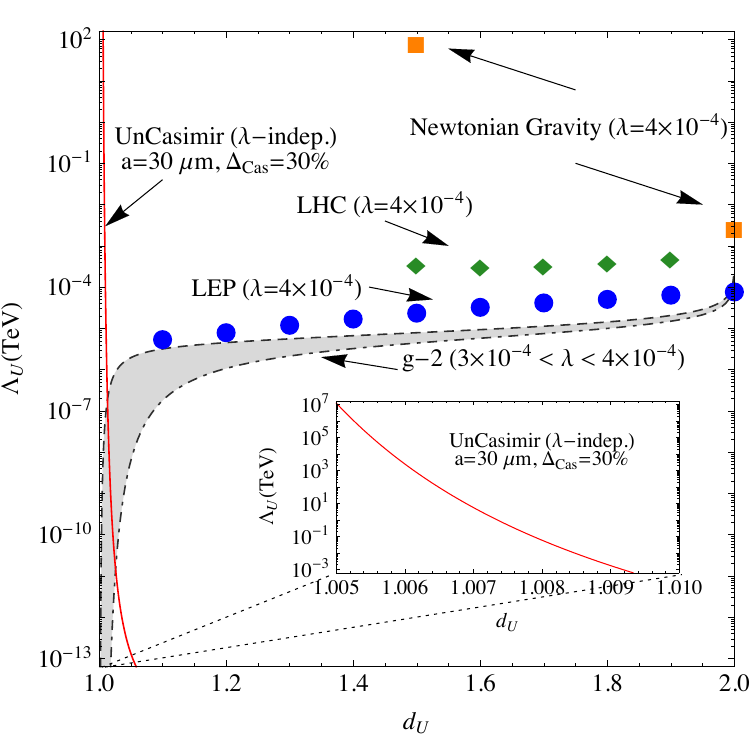} 
\caption{\label{figex1} (Color online) Lower bounds on $\Lambda_{\cal U}$ (the regions below the curves are excluded).
The continuos solid line
is the bound from the Casimir effect~\cite{RevModPhys.81.1827}. 
The central filled area (grey) are the bounds  from the muon anomaly~\cite{PhysRevLett.99.051803}  with different choices of the coupling coefficient $\lambda$.
We include  bounds from direct searches at high energy colliders: LEP~\cite{PhysRevD.84.015010}  full dots (blue) and LHC~\cite{CMS-PAS-EXO-11-043} full diamonds (green).   
The two (orange) square points  are bounds  from sub-millimetre Newtonian gravity~\cite{Goldberg:2007tt,Alderberger:2007aa}.
In the inset we show details of the region $d \approx 1$, where the uncasimir bound is by far the strongest.
}
\end{center}

\end{figure}

In the case of the unparticle contribution to the muon anomaly~\cite{PhysRevLett.99.051803} one has the bound:
\begin{equation}
\label{unparticleg-2}
\Lambda_{\cal U} \ge \Lambda_\mu\!\equiv\! m_\mu \!\left|\frac{\lambda^2 Z_{d_{\cal U}}}{4\pi^2 \Delta_\mu}
\frac{\Gamma(3-d_{\cal U})\Gamma(2d_{\cal U}-1)}{\Gamma(2+d_{\cal U})} \right|^\frac{1}{2d_{\cal U}-2}\!\!\!,
\end{equation} 
where $\Delta_\mu$ is the difference of the experimental result with the Standard Model prediction~\cite{PhysRevD.86.010001}, $\Delta_\mu= \Delta a_\mu (\text{exp}) -\Delta a_\mu (\text{SM}) = 22\times 10^{-10}$ and $Z_{d_{\cal U}} = A_{d_{\cal U}}/(2\sin(\pi d_{\cal U})$. 
We see the above bound is set by the muon mass $m_\mu\approx 105.7$ MeV/$c^2$.   Conversely, the bound  \eqref{bound} is set by the parameter $a^{-1} \approx 2\times 10^{-7}$ MeV/($\hbar c$)  for $a\approx 1 \mu$m.
From \eqref{unparticleg-2}   we see that the g-2 bound depends on the coupling coefficient $\lambda$. 

However we ignore the actual value of $\lambda$. Given the scale hierarchy $\Lambda_{\cal U} < M_{\cal U}<M_\mathrm{Pl}$, the coupling might be smaller, with consequent decrease in predictivity of the muon anomaly analysis. This is not the case  for the un-Casimir effect.  
The lower bound on $\Lambda_{\cal U}$  derived from the Casimir effect  does not depend on $\lambda$ (\textit{cf.}~\eqref{bound}). 
This is in marked contrast with all the proposed bounds in the literature, such as the aforementioned muon anomaly \cite{PhysRevLett.99.051803,Liao:2007bx}, a variety of other particle physics phenomena \cite{PhysRevD.76.075004,Li:2007by,PhysRevD.78.096006}, the predicted deviation of Newton's law at short scales \cite{Goldberg:2007tt}, other astrophysical bounds \cite{Davoudiasl:2007jr,Mureika:2009ii} as well as bounds from atomic parity 
violation~\cite{Bhattacharyya:2007pi}.
This peculiar feature of the un-Casimir effect, \textit{i.e.} being independent of the dimensionless coupling $\lambda $ should not mislead the reader. Like any other physical effect based on the  interaction in \eqref{basic-interaction}, the un-Casimir effect  in reality decouples in the $\lambda \to 0$ limit. 
Indeed, the standard Casimir  formula  for a scalar field of mass $m$, \emph{cf}.  \eqref{ECmassm}, is obtained in the limit {of perfectly conducting plates $\omega_{\text{pl}} a \gg c $ with $\omega_\mathrm{pl}$ the plasma frequency of the conductor.    
In QED, this is equivalent, upon squaring, to 
$\alpha_{\mathrm{EM}}\gg \gamma$, where $\alpha_\mathrm{EM}=e^2/(\hbar c)$ is  the fine structure constant   and $\gamma \equiv\frac{c^2\alpha_{\mathrm{EM}}}{(\omega_\mathrm{pl} a)^2}$ 
a  material dependent quantity scaling as $a^{-2}$.
As noticed in \cite{PhysRevD.72.021301}, Casimir energies are independent of the nature of the plates as well as of any particular interaction coupling ($\alpha_\mathrm{EM}$)
when Dirichlet boundary conditions are perfectly met on metallic plates. In general this is not the case and deviations from the standard Casimir formula increase as the plate separation $a$ decreases. In QED, the deviations become relevant when
$\gamma\sim \alpha_\mathrm{EM}$.
Good conductors~\cite{Rakic:98} (Al, Au, Cu) with $a$ ranging in $[1-50]\,\mu$m have  $\gamma$ from $10^{-5}$ to $10^{-10}$. Therefore, being $\alpha_\mathrm{EM}\approx 1/137\gg \gamma $, one can safely employ the perfect conductor approximation. This is also the case in the un-Casimir effect for 
$\lambda^2/\gamma \gg1  $  and accordingly one finds  $\lambda$-independent {lower bounds} for $\Lambda_{\cal U}$. 
}

The above line of reasoning  is confirmed by the background field approach, \emph{i.e.} the calculation of the Casimir effect by  computing the one-loop effective action due to an interaction ${\cal L}_{\text{int}}=\frac{1}{2}\,g\, \sigma\, \phi^2$ with a sharp background field $\sigma(z)=\delta(z-a/2)+\delta(z+a/2)$. Here the field $\sigma(z)$  mimics the geometrical Dirichlet boundary conditions on the 
plates at a distance $a$ along  $z$-axis~\cite{Graham:2003aa,Graham:2003ib}. Then  the resulting renormalised energy ${\cal E}^C(m,g)\equiv {\cal E}^C(m,g,a)-{\cal E}^C(m,g, a\to \infty)$ reads:
\begin{equation}
\label{bfa}
{\cal E}^C(m,g)=  \int_m^\infty   \frac{dt\, t}{4\pi^2} \,\sqrt{t^2-m^2} \log\left[1- \frac{g^2\,e^{-2at}}{4t^2+4tg+g^2}\right].
\end{equation}
The above relation readily interpolates between the known Casimir result of \eqref{ECmassm} (for $g\to \infty$) and the decoupling limit (${\cal E}_C \to 0$ when $g\to 0$). 
{In QED the coupling $g$ ($dim[g]=E$) is identified with the plasma frequency $\omega_\mathrm{pl}$, 
which is indeed infinite {for} a perfect conductor and therefore identifies with  the strong coupling limit $g\to \infty$. On the other hand the {decoupling takes place in the} limit $g\to 0$  because $g$ scales like $e^4$ (or $\alpha^2_{\mathrm{EM}}$) and therefore vanishes in the limit ($e\to 0$)~\cite{PhysRevD.72.021301}}. 
Our working hypothesis in \eqref{basic-interaction} implies that the unparticle coupling $\lambda$ plays the role of the charge $e$ in QED and any given material will have an unparticle plasma frequency $\widetilde{\omega}_{\text{pl}}$ 
related to the vacuum polarisation fermion loop diagram identical to that of QED with $e\leftrightarrow \lambda$~\cite{A.-A.-Abrikosov:1965qy}. At  one loop order one can therefore calculate the unparicle plasma frequency
\begin{equation}
    \widetilde{\omega}_{\text{pl}}^2 = \left[ \left(\frac{\lambda}{e}\right)^2 \,\frac{\omega_{\text{pl}}^2}{\Lambda_{\cal U}^{2\left(d_{\cal U}-1\right)}} \right]^{\frac{1}{2-d_{\cal U}}}
\end{equation}
that reduces to 
$\widetilde{\omega}_{\text{pl}} \simeq (\lambda/e)\,\omega_{\text{pl}}$ for  $d_{\cal U} \approx 1$ when the uncasimir bound is stronger. 
The perfect conductor approximation for the unparticle Casimir effect is then 
$\widetilde{\omega}_{\text{pl}} \gg c/a $ or $\lambda^2/\gamma \gg1$.
\begin{figure}[t!]
\begin{center}
\includegraphics[width=0.45\textwidth]{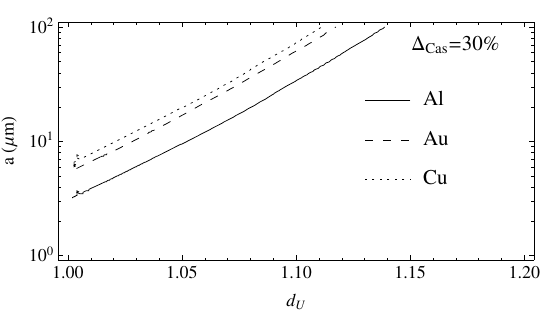} 
\caption{\label{figex3}Contour plot of  the ratio $\Lambda_{a}/\Lambda_\mu=1$ in the decoupling regime. The regions above  the curves correspond to values of the ratio $\Lambda_{a}/\Lambda_{{\mu}}>1$ (un-Casimir provides the strongest bound).}
\end{center}

\end{figure}
However, 
corrections in powers of $1/g$ (i.e. of $1/\widetilde{\omega}_{\text{pl}}$) can be computed  in order to go beyond the perfect conductor approximation.
The first order correction, to eq.~(7) reads:
\begin{equation}
{\cal E}^{C,\,{\cal U}}_{(1)} = \left(\frac{2}{ga}\right) \frac{1}{a^3} \,\frac{ d_{\cal U} \, \zeta(2+d_{\cal U})}{(4\pi)^{2d_{\cal U}}(\Lambda_{\cal U} a)^{2d_{\cal U} -2}}\, =\,-\, \left(\frac{2}{ga}\right) \, {\cal E}^{C,\,{\cal U}}_{(0)}
\end{equation}
where ${\cal E}^{C,\,{\cal U}}_{(0)}$ is the perfect conductor result of Eq.~(7). 
The relative magnitude of the first order correction with respect of the perfect conductor result is ${\cal O}\left({1}/{(ga)}\right)$. This provides us with a physical basis to decide for what numerical values of $\lambda$  the condition $ga \gg 1$ (equivalent to $\lambda^2/\gamma \gg1$) is satisfied. Indeed we can safely apply  the perfect conductor result if the first order correction  is within the experimental error of the Casimir measurement. Therefore given that $1/(ga)=1/(\widetilde{\omega}_{\text{pl}}a)=\sqrt{\gamma}/\lambda$ we  require that  $\lambda \geqslant 2 \sqrt{\gamma}/\Delta_{\text{Cas}}$. 

\section{Results}

Fig.~\ref{figex2} shows the region in the parameter space $(\lambda, d_{\cal U})$ where the un-Casimir bound on $\Lambda_{\cal U}$ wins over the g-2 bound in the strong coupling limit ($\lambda^2/\gamma\gg 1$). The range of applicability of our $\lambda$-independent result --see the dark filled triangle in Fig.~\ref{figex2}-- can be increased either reducing $\gamma$ (other materials, larger distances and/or modulating  the
effective plasma frequency~\cite{Pendry:1996aa,1674-1056-22-8-087302, 1879101,PIERM10031505,Brand:2007aa}) while higher precision in the Casimir measurement may require to go to higher order in the perturbative expansion based on~\eqref{bfa}. The current state of the art in the Casimir effect for perfectly conducting parallel plates is the measurement reported in~\cite{Bressi:2002aa}. 
The relative error on the Casimir energy 
for plates distance $a=1\,\,  \mu$m 
is then $\Delta_{\text{Cas}}\in [21\% - 33\%] $. However, larger plate
separation experiments ($a\in [5,10]$$\,\mu$m) are currently under investigation~\cite{PhysRevLett.113.240405, 1367-2630-8-10-239}, while distances up to $a\approx 50\,\mu$m have been considered in \cite{PhysRevLett.104.241101}. In this regime lowers values of $\gamma$ (and hence $\lambda$) become accessible and the un-Casimir starts to be extremely competitive.
In Fig.~\ref{figex1} we show the results for $\lambda = 4\times 10^{-4}$ where the un-Casimir wins, see Fig.~\ref{figex2}. For $d_{\cal U}$ in the interval [1.005, 1.007] the  bounds on $\Lambda_{\cal U}$ are respectively in the range $[10^7,10]$ TeV.

For $\lambda^2/\gamma\! \ll\! 1$, the decoupling limit  discussed above corresponds to a $\lambda$-dependent un-Casimir energy. The leading ${\cal O}\left((ga)^2\right)={\cal O}\left(\lambda^2/\gamma\right)$ term is computed form~\eqref{bfa}:
\begin{equation}
{\cal E}^{C,\,{\cal U}} = (ga)^2 \, \frac{2^{-(1+2d_{\cal U})}}{(2\pi)^{2d_{\cal U}} (2d_{\cal U} -1)}\, \frac{1}{a^3}\, \frac{1}{(\Lambda_{\cal U}a)^{2d_{\cal U}-2} }
\end{equation} and the contour plot of the unit ratio of the corresponding bound on $\Lambda_{\cal U}$ (now $\lambda$-dependent) with the g-2 bound is presented in Fig.~\ref{figex3} which shows that there is a sensible region in the plane $(a,d_{\cal U})$ where the un-Casimir wins. Note that this region is instead $\lambda$-independent.

In conclusion we have highlighted regions of the parameter space where the bound on $\Lambda_{\cal U}$ from current Casimir experiments is the strongest amongst the ones available, both in the strong coupling regime ($\lambda^2/\gamma\gg 1$) or in the decoupling regime ($\lambda^2/\gamma\ll 1$). By combining the two regimes one finds that for plate distances of about $a\sim 5\mu$m and larger the un-Casimir bound wins over the other bounds for $d_{\cal U}\sim 1$.

\section*{Acknowledgments} 
This work has been supported by the Helmholtz International Center for FAIR within the framework of the LOEWE program (Landesoffensive zur Entwicklung Wissenschaftlich-\"{O}konomischer Exzellenz) launched by the State of Hesse, by the Helmholtz Research School for Quark Matter Studies (H-QM), by the project ``Evaporation of microscopic black holes'' under the grants NI 1282/2-1 and NI 1282/2-2 of the German Research Foundation (DFG),  and in part by the European Cooperation in Science and Technology (COST) action MP0905 ``Black Holes in a Violent Universe''.  This research was supported in part by Perimeter Institute for Theoretical Physics. Research at Perimeter Institute is supported by the Government of Canada through Industry Canada and by the Province of Ontario through the Ministry of Economic Development and Innovation.
The authors are grateful to J. Mureika for valuable comments.

%
%


\end{document}